\documentclass[10pt]{article}
\usepackage{graphicx}
\usepackage{booktabs} 
\usepackage{enumitem}
\usepackage{multicol}
\usepackage{float}
\usepackage{multirow}
\usepackage{tabularx}
\usepackage{etoolbox}
\usepackage[section]{placeins}
\usepackage[utf8]{inputenc}
\usepackage{cite}
\usepackage{amsmath}
\begin{document}
\title{Computation of the zero temperature RSB parameter in Bethe Lattice Spin Glasses}

\date{}

\author{Francesco De Santis\\
			\small Dipartimento di Fisica, Sapienza Universit\'a di Roma,\\
			\small francesco.desantis@roma1.infn.it
           \and
           Giorgio Parisi\\
             \small Dipartimento di Fisica, Sapienza Universit\'a di Roma,\\
             \small Nanotec, Consiglio Nazionale delle Ricerche, I-00185 Rome, Italy \\
			 \small giorgio.parisi@roma1.infn.it
}

\maketitle
\begin{abstract}
    Bethe lattice Spin Glasses (BLSG) are a class of models with finite connectivity, which undergo a Replica Symmetry Breaking (RSB) phase transition at zero temperature, at a critical value of the external field.
   
    We compute numerically the RSB order parameter near the transition, in the case of minimum connectivity ($z=3$), with bimodal distribution of the couplings ($J=\pm1$), and Gaussian external field. We use an approach based on a probabilistic computation which does not require the knowledge of the RSB scheme. The method exploits a universal formula which relates the zero temperature order parameter to the joint probability distribution of the energy difference and overlap of excited states induced by a convenient perturbation to the Hamiltonian.
   
\end{abstract}

\section{Introduction}
\label{intro}
Spin glasses differ from normal magnets because of the non-trivial structure of their configuration space, characterized by a rough free-energy landscape and a large number of equilibrium states.
This glassy behavior is generally well understood in the case of infinite range models, where the mean-field approach provides exact solutions.

In this case, the equilibrium states are organized in the phase space according to a non-trivial structure, in a scenario known as Replica Symmetry Breaking (RSB) \cite{MEZPARVIRSGAB}. The complexity of this description is captured by the spin glass order parameter, which is identified with the probability distribution of the overlap between the states. In systems characterized by \emph{full} RSB, the order parameter is typically a continuous function of the overlap.
However, the validity of this scenario in short range models is still debated, and determining the spin glass order parameter could be crucial to understanding to which extent the RSB picture describes their low-temperature phase.

Studying the systems directly at zero temperature has some advantages. In the first place, pure states are simple configurations, and the statistical properties of the ensemble can be studied in terms of the ground state energy, instead of the free-energy. In general, there is no proof that the description at zero temperature and finite temperature are the same, but it is believed \cite{MARPARETAL2000} that if, under certain assumptions, the zero temperature limit and the thermodynamic limit commute, then the statistical properties at zero temperature can be extended to the finite temperature region. Another advantage comes from the availability of efficient algorithms that work at zero temperature.\\

The investigation into the consequences of RSB can be done by studying the space of the low-lying excitations induced by a bulk perturbation, as suggested in \cite{KRZMAR2000, PALYOU2000,MARPAR2001}. 
The method, known as $\epsilon$-coupling, consists in introducing a repulsive coupling between the system and its ground state and studying how the energy of the low-lying configurations varies as a function of the overlap. In \cite{FRAPAR2000} a formula is found for the non-trivial joint distribution of energy gap and overlap induced by the perturbation, in terms of the order parameter. 
We used this relation to compute the order parameter at zero temperature of the Bethe lattice spin glass (BLSG) with bimodal couplings, and study the RSB transition driven by the external magnetic field.\\

The reason why we are interested in this model is twofold.
First, from the theoretical physics point of view, it is a particular case of a model with finite connectivity with a topology which facilitates the analytic approach to its full RSB solution.
Nevertheless, the exact solution has not been found yet.
An approximated solution is provided by the cavity method \cite{MEZPAR2001, MEZPAR2003}, at a level equivalent to the first step of replica symmetry breaking (1RSB).
In this respect, we believe the method we apply in our study to be a valid tool of investigation for either supporting the theoretical advance towards the full RSB solution or for understanding to what extent the system differs from the RSB phase.

Second, from the computational point of view, the particular topology of the BLSG suggests that a particular kind of algorithms based on the iterative optimization of large portions of the graph might be particularly effective \cite{HAR1996}, facilitating the computation of exact ground states.\\
 
In the first section, we shall define the RSB parameter and outline the derivation of the formulas which will be used (Sec. \ref{sec:RSB}). We will continue with a brief description of the model (Sec. \ref{sec:BLSG}). Thus, we shall explain the numerical approach (Sec. \ref{sec:CEA}) and present the results (Sec. \ref{sec:Results}), followed by some conclusions and perspectives of our work (Sec. \ref{sec:end}).

\section{The RSB order parameter}
\label{sec:RSB}
The physical spin glass order parameter is the probability distribution $P(q)$ of the mutual overlaps $q$ among the large number of states\cite{MEZPARVIRSGAB}.
If $\alpha, \beta$ indicate two different states, the overlap and its probability distribution are defined:
\begin{equation}
    q_{\alpha \beta}=\frac{\sum_i m_i^\alpha m_i^\beta}{N} \, ,    \quad P(q) =     \sum_{\alpha \beta} \delta (q - q_{\alpha \beta} ) w_\alpha w_\beta
\end{equation}
where $w_\alpha$, $w_\beta$ are the weights associated with the Boltzmann decomposition.
It is convenient to define the integrated probability function
\begin{equation}
    x(q) = \int_{-1}^q P(q')\; dq'
\end{equation}
whose inverse function $q(x)$ corresponds to the $Q(x)$ order parameter originally defined in the replica formalism.

In the replica symmetric (RS) phase, $P(q)$ is expected to be a single delta function centered on a finite value of overlap. Upon a transition to the full RSB phase, $P(q)$ becomes a continuous function in a finite interval.

One of the most important consequences of RSB is the validity of special \emph{ultrametric} relations\cite{MEZPARETAL1984}. Ultrametricity implies that the probability that the overlaps of three states chosen at random are arranged in a shape different from equilateral or isosceles triangles, is zero. In mean-field models, this is a direct consequence of the hierarchical Ansatz chosen for breaking the replica symmetry. The resulting structure can be represented on a hierarchical tree in which the states (at the bottom of the tree) are clustered at several levels, so that states branching from the same node at a certain level $k$ have overlap $q>q_k$. The \emph{full RSB} scenario corresponds to the limit of continuous branching.\\

As mentioned, we are interested in determining the order parameter at zero temperature. In this case, the pure states reduce to single configurations. We assume that $P(q,T)$ and $x(q,T)$ are smooth functions of the temperature close to $T=0$ \cite{PARTOU1980,MARPARETAL2000}:
\begin{equation}
    x (q,T) = T y(q) + O(T^2)
\end{equation}
where $y(q)$ may be singular at $q=1$.

Although the approach we are going to describe is very general, and no model has been defined yet, we should remark that we will consider an external magnetic field. This has two effects on the system. First, it breaks the global spin inversion symmetry, and therefore the order parameter is expected to be defined only for non-negative values of overlap. Second, it breaks the degeneracy of the ground state, which is unique. 
\\

We follow the method suggested by Franz and Parisi \cite{FRAPAR2000} for computing the RSB order parameter of the BLSG.
The idea consists in perturbing the Hamiltonian with a repulsive term between the system and its ground state.
If the perturbation is order $O(1)$, the distribution of the overlaps between the unperturbed ground state and the one in presence of the perturbation, is non-trivial. The spin glass order parameter is thus related to the overlap distribution of the low-lying excitations through a formula, valid in cases in which ultrametricity holds.
We exploited this relation for extracting numerically the RSB order parameter.
In this section, we outline the derivation of the formula, skipping the details, which can be found in \cite{FRAPAR2000}.\\

Let us consider the Hamiltonian:
\begin{equation}
   \label{eq:Heps}
   \mathcal{H}_\epsilon(\sigma) = \mathcal{H}_0(\sigma)+ \epsilon q (\sigma)
\end{equation}
where $\mathcal{H}_0$ is the original Hamiltonian, $\sigma_0$ is the unperturbed ground state, and $q(\sigma)=\sum_i \sigma^i \sigma_0^i$ is the extensive overlap between $\sigma_0$ and a generic configuration $\sigma$.
The ground state of $\mathcal{H}_\epsilon$ is an excited state for $\mathcal{H}_0$ with energy:
\begin{equation}
    \displaystyle E_{GS}(\epsilon) =  E_{GS}(0)  + \min_{0\le q \le 1}\{\Delta(q) + \epsilon q\}
\end{equation}
Let $q$ be the value which minimizes the rhs of the previous formula, and let $\Delta$ be the relative energy difference.
As $0\le q\le1$, the energy gap $\Delta$ verifies the inequality $0 \le \Delta < \epsilon (1-q)$, being zero in the case $q=1$.\\

The analytic expression of the related joint probability distribution $P(\Delta,q)$, can be obtained in terms of the order parameter by means of a probabilistic computation. The derivation is completely general for systems in the RSB phase, and it is based on two assumptions: (i) an exponential distribution of the energies and  (ii) ultrametricity \cite{MEZPARETAL1984,MEZVIR1985}.
The computation is done by considering an arbitrary ultrametric tree, starting from the simple case of 1RSB, iterating the formula in the case of $k$-step, and finally considering the continuum branching limit, which corresponds to the case of full RSB.
One of the main results of \cite{FRAPAR2000} is the following formula:
\begin{equation}\label{eq:Pdq}
    \centering
    \begin{split}
        &P(\Delta,q)= \\
        &\theta (1-q-\Delta/\epsilon) y'(q) \exp \left({ -\epsilon \int_0^{q+\Delta/\epsilon}  y(q') -y(0)\;dq}\right) + \\
        &+ \delta(\Delta) \delta(q-1) \exp\left( -\epsilon \chi \right)
    \end{split}
\end{equation}
where $\chi = y(0) + \int_0^1dq y(q)$, and $\Delta, q$ satisfy the constraint $\Delta < \epsilon (1-q)$.
Equation \ref{eq:Pdq} could be integrated over $\Delta$, obtaining an expression for $P(q)$, depending on both the derivative and the primitive of $y(q)$.
A simpler expression can be obtained noticing that Eq.\ref{eq:Pdq} depends on $\Delta$ only in the combination $w = q +\Delta/\epsilon$. Integrating over $\Delta, q$ for fixed $w$ with the condition $0<q<w$:
\begin{equation}\label{eq:Pw}
    \begin{split}
        P(w) = &\theta(1-w) \left(y(w)- y(0)\right)\epsilon \exp \left(-\epsilon Y(w)\right) + \\
        &+\delta(w-1) \exp(-\chi)\\
    \end{split}
\end{equation}
where 
\begin{equation}\label{eq:yw_def}
    Y(w) \equiv \int_{-\infty}^w\, y(q) - y(0)\, dq
\end{equation}
A simple expression can be found by considering:
\begin{equation}\label{eq:qw}
    Q(w) \equiv \int_w^1 P(w')\; dq \; = \exp -(\epsilon Y(w))
\end{equation}
In the end the following expression is found:
\begin{equation}\label{eq:yw}
 Y(w) = - \frac{\log Q(w)}{\epsilon}
\end{equation}
This formula is very useful for computing $y(q)$ numerically, as $Y(w)$ is related to the order parameter by Eq.\ref{eq:yw_def} and $Q(w)$ can be extracted in simulations.

We do not have an analytic expression of $y(q)$ for the BLSG. We know that in the SK model $y(q)\ne 0$, and has form $y(q)\sim q (1-q)^{-1/2}$ \cite{MARPARETAL2000}.\\

Hereafter we shall refer to $Y(w)$ as the RSB order parameter, as it is statistically more stable than $y(q)$, and therefore more suited for numerical estimation.

\section{The model}
\label{sec:BLSG}
We define our spin glass model on a graph $G$ with $N$ vertices, and $M$ edges. A spin variable $\sigma_i$ is assigned to each vertex $i$, and a coupling coefficient $J_{ij}$ representing a pairwise interaction is assigned to each edge $(i,j)$. The coefficients $J_{ij}$ are quenched, independent and identically distributed random variables sampled from a probability distribution $P(J)$. 

In general, a Bethe lattice is a graph where the Bethe-Peierls approximation is exact, due to the possibility to neglect the correlations introduced by closed paths in the lattice. In spin glasses, where closed paths are needed for implementing frustration, a sufficient condition for a graph to be a Bethe lattice is that the density of loops of finite size goes to zero in the thermodynamic limit. In this limit, the graph is locally a tree.

In this paper, we consider spin glasses defined on Random Regular Graphs (RRG), a particular class of random graphs in which every node has a fixed number $z$ of neighbors. This class of graphs has a homogeneous topology which is locally tree-like in the thermodynamic limit and it is therefore a good representation of Bethe lattice.
The number of edges $M=zN$ must be an even number, and the graph does not contain any \emph{self-loop} $(i,i)$ or multiple edges between the same nodes.

We focus on spin glasses with Ising spin variables and bimodal couplings $J=\pm 1$, focusing on the minimum value of connectivity $z=3$. Moreover, we introduce an external magnetic field $h$ to drive the RSB transition, sampled from a Gaussian distribution with zero mean and variance $H^2$.

The Hamiltonian of the model is:
\begin{equation}
    \mathcal{H} = -\sum_{<i,j>} J_{ij} \sigma_i \sigma_j - \sum_{i=1}^N h_i \sigma_i
\end{equation}

The BLSG has been widely studied analytically, both at finite and zero temperature \cite{MEZPAR2001, MEZPAR2003}.
The \emph{cavity} method, a generalization of the Bethe approximation, shows the presence of an instability of the RS solution along the \emph{de Almeida-Thouless} (AT) line in the $(H,T)$ plane, and provides a solution equivalent to the 1-step of RSB in the unstable region.
At this level of approximation, many thermodynamic quantities show a good agreement with the numerical experiments. However, a further investigation into the nature of the RSB phase is not possible without pushing the calculation to further orders.

In the $z\to \infty$ limit, the BLSG is expected to reduce to the SK model. The high connectivity limit has been studied in terms of the $1/z$ expansion\cite{DEDOMGOL1989, PARTRI2002EPJB} at the 2RSB level, and there is evidence that the discrepancy between the BLSG and the SK for $z\to \infty$ vanishes when the computation is brought to further levels of RSB.

However, the behavior at low temperature is rather different at finite $z$. In fact, while the SK model is full RSB and exhibits no transition at zero temperature, due to the divergence of the AT line for $T \rightarrow 0$, in models with finite connectivity the critical line converges to a finite value $H_C$ in the zero temperature limit. As a result, the BLSG undergoes an RSB transition at the critical value $H_C$. A method for obtaining the AT line at low temperature is described in \cite{PAGPARRAT2003}, and is based on the study of the cavity equations of the $\epsilon$-coupled Hamiltonian introduced in the previous section.

For the BLSG with connectivity $z=3$ and $J\pm$ interactions the critical value at zero temperature is $H_C\sim1.037$ \cite{PC01}.\\

Several numerical results on BLSG at zero temperature have been published in the past\cite{BOE2003EPJB, BOE2003PRB, BOE2010JSM}, mostly focusing on the behavior in zero external magnetic field.
They report that, despite the BLSG model should tend to the SK model in the large $z$ limit, the behavior at finite $z$ depends on the particular choice of the $P(J)$.
For instance, while in the $J\pm$ model the finite-size corrections to the ground state energy scale as $N^{-\omega}$ with $\omega = 2/3$, like in the SK model, in the Gaussian model they seem to scale with a different value of the exponent $\omega=4/5$, unless higher order corrections are taken into account. \cite{BOE2010JSM}.

Besides obtaining the RSB order parameter at zero temperature, we also aim to compute the finite-size corrections to the ground state energy in the presence of a finite magnetic field.

\section{Numerical approach}
\label{sec:CEA}

The numerical analysis requires the computation of exact ground states. In the case of spin glasses, this is a typical hard optimization problem, mostly because of the presence of frustrated loops. In this respect, the efficiency of an algorithm might depend on the particular topology of the lattice.

In Bethe lattice spin glasses, short loops are expected to be rare for large $N$, and frustration occurs only on large scale. This suggests that the solution could be achieved by optimizing iteratively large portions of the graph (clusters) free of frustration, until a minimum is eventually reached.

Very efficient techniques for optimizing the clusters include \emph{Belief Propagation} (BP) \cite{MEZMONIPC}, a powerful message-passing algorithm which can be applied to trees, and MINCUT techniques, which can be also used on graphs.
In these cases, the capacities of the edges (coupling interactions) need to be non-negative.
The BP version of the algorithm was implemented by Decelle and Krzakala \cite{DECKRZ2014}, and can also be used at finite temperature. The algorithm using MINCUT was implemented by Hartmann \cite{HARUSA1995, HAR1996, HAR1999} and is known as \emph{Cluster Exact Approximation} (CEA).

Despite MINCUT being computationally more expensive on large graphs than BP, we preferred to use the CEA, due to the possibility of including loops in the selected clusters. The problem of the regularization of negative coupling can be solved by applying a local gauge transformation, provided the loops are not frustrated.\\

We first generated a large number of samples $({\sim10^4} - 10^5)$ of different size $N = 200, 400, 800,$ $1600, 3200$, with random coupling interactions $J=\pm1$, and random Gaussian field, for different values of variance $H^2$, with $H=0.6, 0.8,$ $1.0,1.2$.\\

The ground state of each graph was computed by means of the CEA. This algorithm consists of two phases: \emph{clustering} and \emph{optimization}. During the clustering phase, a subgraph is generated starting from a random root and adding iteratively new nodes only if their inclusion does not introduce frustration. In this way, it is possible to define a gauge transformation over the subgraph, as mentioned before.
During the second phase, the cluster is optimized by means of a MINCUT technique (our version was based on an incremental breadth-first search algorithm \cite{ALG01, ALG02}). The output of the algorithm is the configuration (cut) of the subgraph variables which minimizes the number of unsatisfied bonds, with respect to the boundary.
The CEA dynamics is thus characterized by a monotonically decreasing energy, whose minimum is eventually reached by iterating the clustering-optimization procedure for a fixed number of steps $n$.

Despite the ability of the CEA to avoid local traps that would invalidate a standard zero temperature Monte-Carlo algorithm based on Metropolis dynamics, a single run of the algorithm is not sufficient to find an exact ground state. Therefore, we applied the algorithm multiple times starting from different random configurations, until the same ground state was found ten times.

The typical size of the clusters generated by the CEA is $\sim 0.8$ of the whole graph for $z=3$, and it seems to be independent of $N$, in our range of values. The algorithmic complexity is $O(N^\alpha)$, with $\alpha \sim 3$. As the number of starting conditions needed for finding the exact ground state depends on $n$, it is important to tune this parameter before running the simulations.

\section{Results}
\subsection{Computation of $Y(w)$}
\label{sec:Results}
We considered two different class of systems, both defined on Bethe lattice:
\begin{enumerate}
    \item Ising spin glasses with random $J=\pm1$;
    \item Ising ferromagnetic systems with $J=1$;
\end{enumerate}
In both cases, we applied an external Gaussian field with variance $H^2$.
In the first case, we expect to observe replica symmetry breaking at $H<H_C \sim 1.027$. The second case, instead, corresponds to a Random Field Ising Model (RFIM) on an RRG, where no RSB phase is expected. In these systems there is a transition from the paramagnetic to the ferromagnetic phase at $H=1$.\\

\begin{figure*}
    \centering
    \includegraphics[width=.95\textwidth]{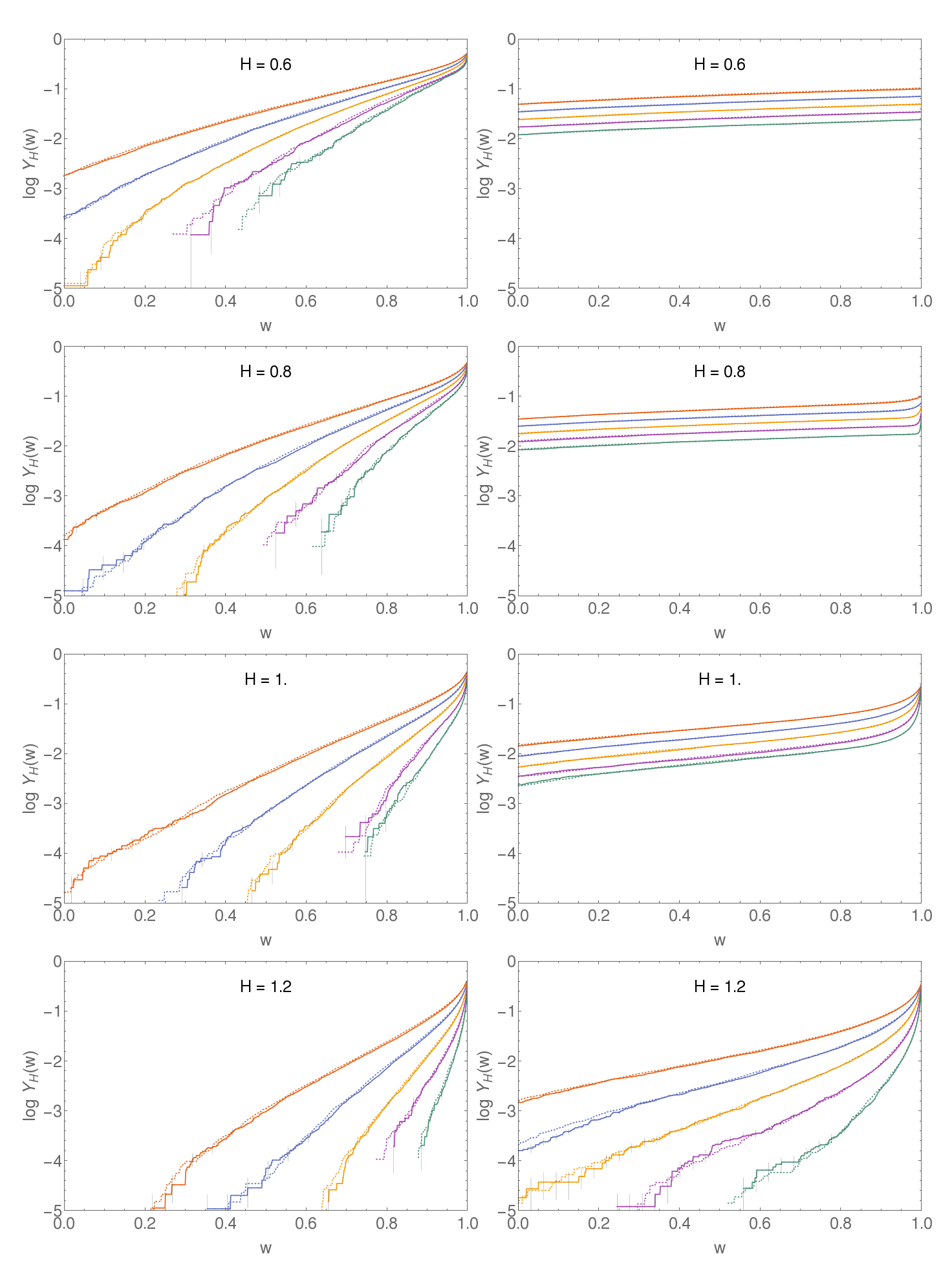}
    \caption{Overlap distribution $Y_{H,N}(w)$ of the BLSG (left column) and RFIM (right column), for different values of field variance $H^2$. In each box, every line lines represent systems of different size ($N=200, 400, 800, 1600, 3200$, from top to bottom). Different values of $\epsilon$ are also represented: $\epsilon = 1$ (solid)  and $\epsilon=2$ (dashed).}
    \label{fig:ywn}
\end{figure*}
  
\begin{figure*}
    \centering
    \includegraphics[width=.95\textwidth]{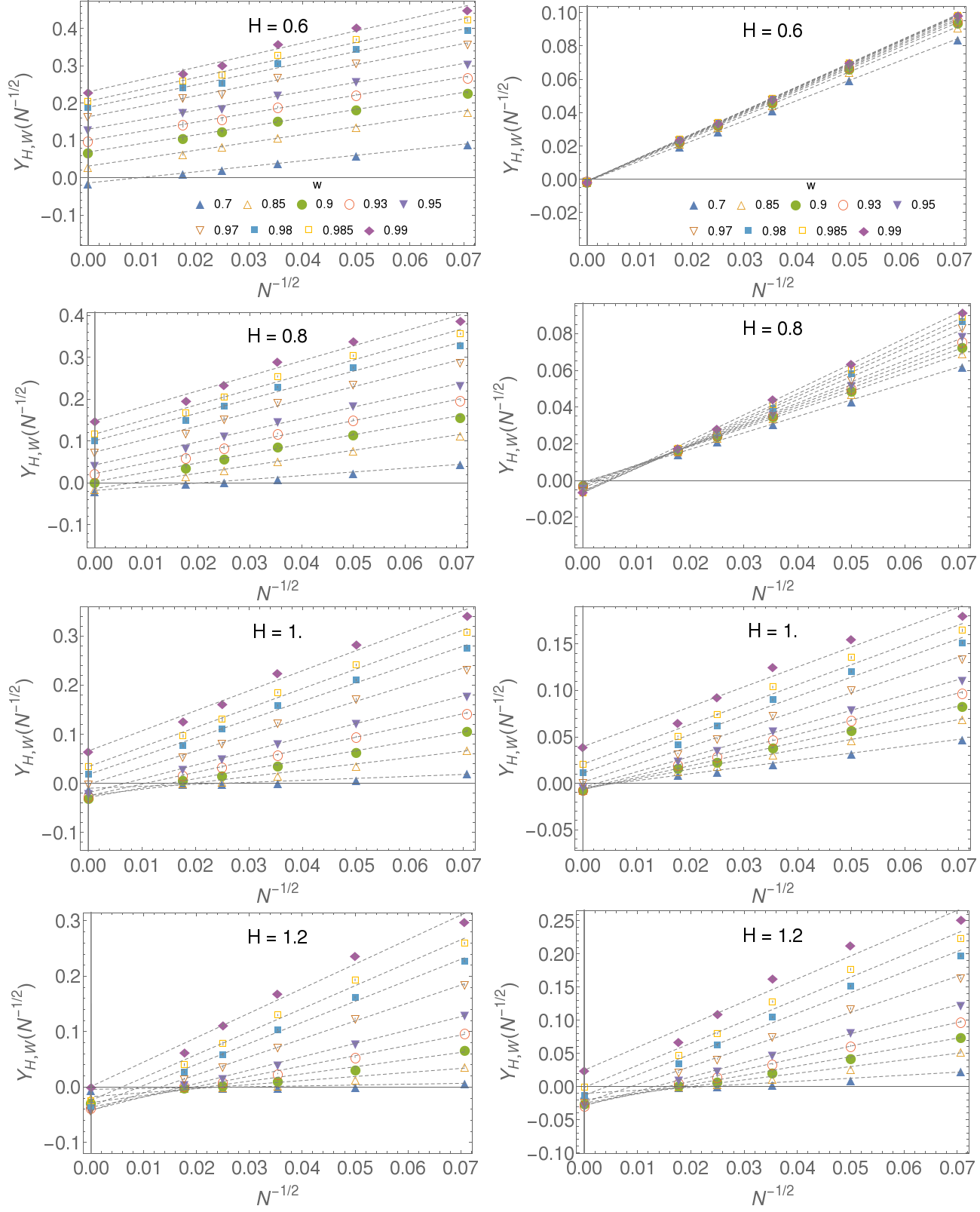}
    \caption{Extrapolation plot for the function $Y_H(w)$. The data is plotted at fixed values of $w$ versus $N^{-1/2}$, according to Eq. \eqref{EQ_REGRESSION}. The quality of the fit decreases for $w\to 1$. The points on the negative y-intercept follow a different scaling form and extrapolate to zero. }\label{fig:linreg}
\end{figure*}

\begin{figure*}[t]
    \centering
    \includegraphics[width=.85\textwidth]{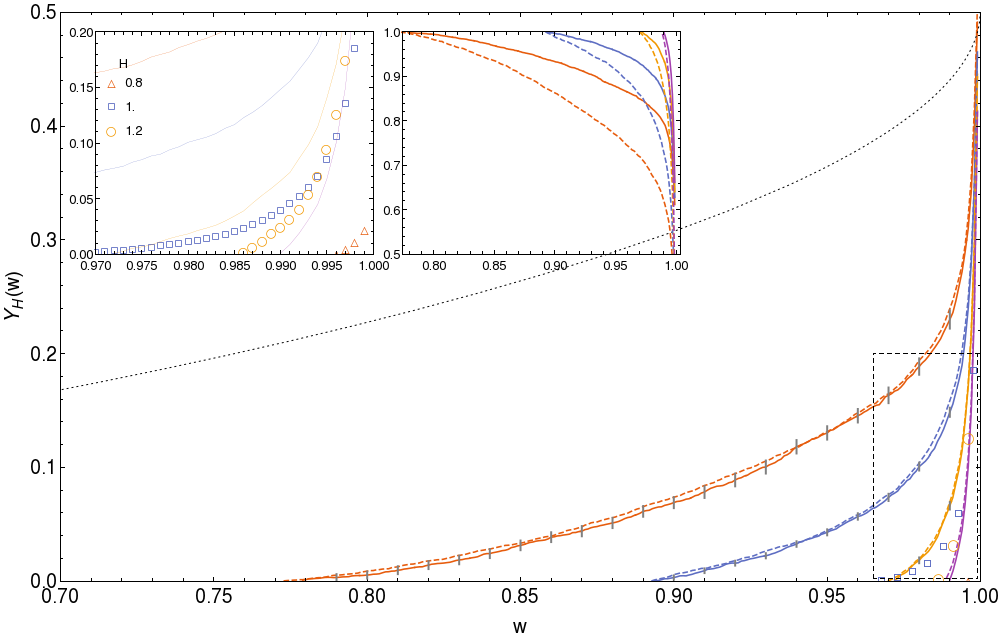}\\
    \caption{ $Y_H(w)$ in the large $N$ limit, in BLSG (lines), and RFIM (points), for different values of $H=0.6,0.8,1.0,1.2$ (from top to bottom), for $\epsilon=1$ (solid lines) and $\epsilon=2$ (dashed lines). The dotted isolated line represents $Y(w)$ for the SK model (using the form $y(q) = (aq + bq^2)(1-q)^{-1/2}$ as suggested in \cite{FRAPAR2000}). The left inset is a zoom on the small dashed region, where the RFIM points are concentrated. In the right inset, the $Q_H(w)$ distributions are plot for increasing values of $H$ (from bottom to top), and different values of $\epsilon$ (solid and dashed lines). It is evident that $Y(w)$ does not depend on $\epsilon$, differently from $Q(w)$.}\label{fig:yw}
\end{figure*}

    For each sample we obtained the ground state $\sigma^0$ of the unperturbed Hamiltonian $\mathcal{H}_0$, with energy $E_{GS}(0)$, and the ground state $\sigma^\epsilon$ of $\mathcal{H}_\epsilon$ defined in Eq.\ref{eq:Heps}, with energy $E_{GS}(\epsilon)$. The analysis was done for two values of $\epsilon =1, 2$.
    
    From the simulations, we obtained the energy difference $\Delta(\epsilon)=E_{GS}(\epsilon)-E_{GS}(0)$, the overlap between the two ground states $q$, and the relevant quantity $w = q + \Delta/\epsilon$, whose distribution is related to the order parameter.
    We should point out that the analysis which follows is based on the assumption that the algorithm succeeds at finding exact ground states. In order to reduce the sources of error we filtered the data by discarding the samples with (a) negative values of $\Delta$, which indicate a wrong identification of $\sigma^0$, and (b) $w>1$, corresponding to samples in which $\sigma^\epsilon$ has energy $E(\epsilon)>E_{GS}(0)+\epsilon$, which cannot correspond to a ground state. The amount of data we discard is less than the $1\%$ of the whole sample, the number of occurrences slightly increasing in larger systems.\\
    
    In order to estimate the error associated with the distributions, we used jackknife resampling. We generated 1000 resampled data sets by extracting half of the observations at random from the original data set. For every sample, $Y_{H,N}(w)$ is computed using Eq. \ref{eq:qw}-\ref{eq:yw}, and then averaged over the samples. The results are shown in Fig.\ref{fig:ywn}.\\

The $Y_{H,N}(w)$ curves are increasing functions of $w$, singular in $w=1$. For small values of $N$, negative values of $w$ are observed in the RSB phase. This is not a problem at a finite size, where ultrametricity does not hold strictly, but we expect that in the thermodynamic limit the functions are null for $w<0$. The curves show only a weak dependence on $\epsilon$, supporting the assumptions that $Y(w)$ should not depend on $\epsilon$, as long as it is of order $O(1)$, due to Eq. \ref{eq:yw}.\\

The data seem to scale linearly when they are plotted for $N^{-\alpha}$ with $\alpha = 1/2$, and the curves for $N \to \infty$ were extrapolated by considering the asymptotic form:
\begin{equation}
    \label{EQ_REGRESSION}
    \qquad Y_{H,N}(w) = Y_H(w) + g_H(w) N^{-1/2}
\end{equation}
The fitted lines are in good agreement with the data for $w$ not too close to $w=1$ (see Fig. \ref{fig:linreg}).
We observed that the curves with $w$ below a critical value $w_c(H)$, tend for $N \to \infty$ to negative values on the y-intercept, which are non-physical. This behavior was interpreted by assuming that the functions scale exponentially for $w<w_c$, rather than with a power-law. The critical values $w_C(H)$ are set to the minimum $w$ whose fitted lines extrapolate to a positive value.

In this way, we extrapolated $Y_H(w)$ in the large $N$ limit, as shown in Fig. \ref{fig:yw}.
The resulting $Y_H(w)$ are  positive continuous and monotonically increasing functions defined in the interval $[w_C(H),1]$, singular in $w=1$.

It is evident that $w_C(H) \to 1$ for $H \to H_C$, although $Y_H(w)$ does not tend completely to a $\delta$ function, even for $H>H_C$. However, the quality of the fit is low near the singularity, and the order parameter could reasonably be a $\delta$ in the RS phase.

In the RFIM case, where no RSB transition is expected, the linear fit is very good in the ferromagnetic region ($H<1$), and $Y(w)$ is a sharp delta. In the paramagnetic region $H>1$, the behavior is similar to the previous model.

We remark that, even if the distributions $Q(w)$ extracted from the data depend on $\epsilon$ (Fig. \ref{fig:yw}, right inset), the order parameter $Y(w)$ does not depend on the perturbation.

\subsection{Finite-size scaling}
\begin{figure}
    \centering
    \includegraphics[width=.45\textwidth]{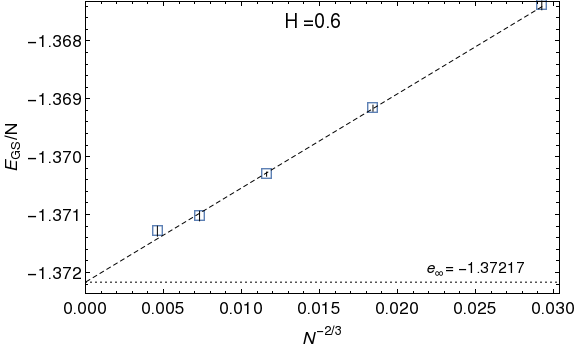}
    \includegraphics[width=.45\textwidth]{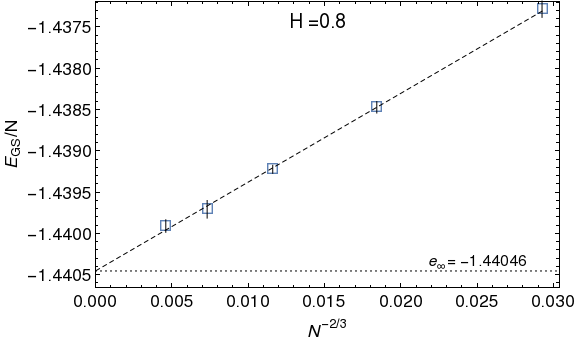}
    \includegraphics[width=.45\textwidth]{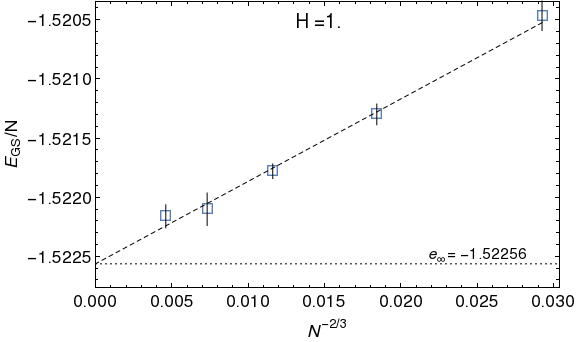}
    \includegraphics[width=.45\textwidth]{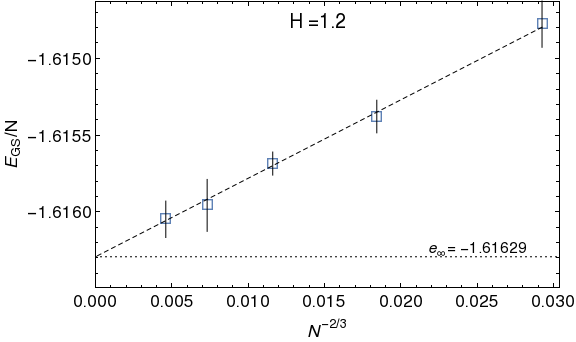}
    \includegraphics[width=.45\textwidth]{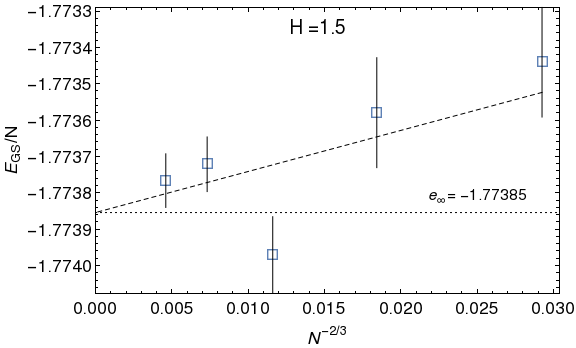}
    \caption{Extrapolation plot for the ground state energies of BLSG. The data is plotted versus $N^{-2/3}$ according to Eq.\ref{eq:Eform}}\label{fig:energyscaling}
\end{figure}

In the second part of the analysis, we computed the finite size corrections to the ground state energy of the BLSG. The energies fit very well to a line when plotted versus $N^{-\omega}$ with $\omega=2/3$ (see Fig. \ref{fig:energyscaling}), for both the RSB and RS phase. The quality of the best fit is worse for higher values of $H$ (e.g. $H=1.5$), due to the higher dispersion.
The ground state energies in the limit $N\to \infty$ are extrapolated by fitting to the form:
\begin{equation}
    \qquad e_{GS}(N)= e_{\infty}+ A N^\omega
\label{eq:Eform}
\end{equation}
where $e_{GS}$ is the ground state energy per spin.
The exponent $\omega=2/3$ is the same measured at zero field by S.Boettcher in \cite{BOE2003EPJB}.
The asymptotic values for the ground state energy per spin are reported in Tab. \ref{tab:Eform}.

\begin{table}[h]
    \centering
    \begin{tabular}{c|lll}
        \hline
        $H$&&$e_\infty$&\\
        \hline
        0.6 & -1.37217& $\pm$ & $6 \cdot 10^{-5}$\\
        0.8 & -1.44046& $\pm$ & $3 \cdot 10^{-5}$\\
        1.0 & -1.52256& $\pm$ & $6 \cdot 10^{-5}$\\
        1.2 & -1.61629& $\pm$ & $3 \cdot 10^{-5}$\\
        1.5 & -1.77388& $\pm$ & $9 \cdot 10^{-5}$\\
        \hline
    \end{tabular}
    \caption{Asymptotic values for the ground state energy in the large $N$ limit, extrapolated using Eq.\ref{eq:Eform}}
    \label{tab:Eform}
\end{table}

\section{Conclusions}
\label{sec:end}

We computed $Y(w)$, a function of the order parameter $y(q)$ (see Eq. \ref{eq:yw}), in presence of Gaussian external field, for two models defined on Bethe lattices: spin glasses with $J\pm1$ and random field Ising models with $J=1$.
In the first case, where a RSB transition is expected to occur at $H=H_C$, $Y(w)$ is a continuous function in the RSB phase, and tends to a delta function $\delta(w-w_C)$ when $H \to H_C$.
In the second case, where no RSB is present, $Y(w)$ is a sharp $\delta$ in the ferromagnetic case, and behaves similarly to the spin glass model in the paramagnetic phase, where the random field is supposed to dominate.

We remark that our analysis relies on the hypothesis that ultrametricity holds. In \cite{FRAPAR2000}, a violation of the scaling form of $Y(w)$ with $\epsilon$, is observed for large values of the perturbation, in systems where ultrametricity does not hold. The fact that in our simulations $Y(w)$ does not depend on $\epsilon$, seems to confirm that ultrametricity holds in the BLSG. A comparison of our results with a future analytic derivation of $y(q)$ might provide information on possible deviations from the full RSB description in our model.

In the BLSG case, we also computed the scaling exponent of the finite-size corrections to the ground state energy below $H_C$, finding $\omega=2/3$. The value is in agreement with previous observations.\\

The method could be extended to larger systems and higher connectivities. However there is a limit imposed by the algorithmic complexity. Despite the CEA being particularly effective on systems with tree-like topology, the number of starting conditions needed for obtaining the exact ground states grows quickly with the size of the systems, and we suspect the algorithm to fail at finding exact ground states for $N>3200$. The performance is expected to be worse for higher connectivity, as short loops are more frequent. Nevertheless, faster algorithms could provide some small improvement.
For instance, a similar Monte-Carlo algorithm based on \emph{Belief Propagation} (BP) is described in\cite{DECKRZ2014} for optimizing spin glasses at finite temperature. However, due to the requirements of BP algorithms, the selected sub-graphs are trees whose size is $\sim 0.7 N$.
In our simulations, we observed that the algorithm performs better when the size of the clusters is larger, as states lower in energy are found much faster and less starting conditions are required.\\

It would be interesting to extend our study to diluted models with spatial structure. A possible extension includes lattices in finite dimensions with where every site interacts with a fixed number of neighbors $z<2D$. Such an analysis might give a perspective on the role of the space dimension on the thermodynamic properties when the number of neighbors is kept constant.
\subsection*{Acknowledgements}
We would like to thank Carlo Lucibello, Aurelien Decelle and Maria Chiara Angelini for a helpful conversation.
We also acknowledge very helpful correspondence with Stefan Boettcher.

This project has received funding from the European Research Council (ERC) under the European Union’s Horizon 2020 research and innovation program (grant agreement No [694925]).

\bibliography{mybibtexshort}{}

\bibliographystyle{unsrt}
\end{document}